\documentclass[%
prd,
reprint,
superscriptaddress,
preprintnumbers,
nofootinbib,
amsmath,amssymb,
aps,
]{revtex4-1}
\pdfoutput=1
\usepackage{hyperref}
\usepackage{graphicx}

\usepackage[textwidth=1.25cm,textsize=footnotesize,
]{todonotes}

\usepackage[utf8]{inputenc}

\usepackage{color}
\usepackage[normalem]{ulem}

\newcommand{\be}{\begin{equation}}
\newcommand{\ee}{\end{equation}}
\newcommand{\ba}{\begin{eqnarray}}
\newcommand{\ea}{\end{eqnarray}}

\def\srg{\emph{SRG}}
\def\erosita{\emph{eROSITA}}
\def\artxc{\emph{ART-XC}}
\def\xmm{\emph{XMM}-Newton}
\def\nustar{\emph{NuSTAR}}

\begin{document}
\title{%
  Towards Testing Sterile Neutrino Dark Matter with Spectrum-Roentgen-Gamma Mission
}

\author{V.\,V.\,Barinov}
\email[]{barinov.vvl@gmail.com}
\affiliation{%
  Physics Department, M. V. Lomonosov Moscow State University, Leninskie Gory, Moscow 119991, Russia}
\affiliation{%
  Institute for Nuclear Research of the Russian Academy of Sciences, Moscow 117312, Russia}
\author{R.\,A.\,Burenin}
\email[]{rodion@hea.iki.rssi.ru}
\affiliation{%
  Space Research Institute of the Russian Academy of Sciences, Moscow 117997, Russia}
\affiliation{%
  National Research University Higher School of Economics, Moscow 101000, Russia}
\author{D.\,S.\,Gorbunov}
\email[]{gorby@ms2.inr.ac.ru}
\affiliation{%
  Institute for Nuclear Research of the Russian Academy of Sciences, Moscow 117312, Russia}
\affiliation{%
  Moscow Institute of Physics and Technology, Dolgoprudny 141700, Russia}
\author{R.\,A.\,Krivonos}
\email[]{krivonos@cosmos.ru}
\affiliation{%
  Space Research Institute of the Russian Academy of Sciences, Moscow 117997, Russia}

\bigskip
\preprint{INR-TH-2020-032}

\begin{abstract}
  We investigate the prospects of the \srg\ mission in searches for
  the keV-scale mass sterile neutrino dark matter radiatively decaying
  into active neutrino and photon. The ongoing all-sky X-ray survey of
  the \srg\ space observatory with data acquired by the \artxc\ and
  \erosita\ telescopes can provide a possibility to fully explore
  the resonant production mechanism of the dark matter sterile
  neutrino, which exploits the lepton asymmetry in the primordial
  plasma consistent with cosmological limits from the Big Bang
  Nucleosynthesis. In particular, it is shown that at the end of the
  four year all-sky survey, the sensitivity of  the \erosita\
  telescope near the 3.5 keV line signal reported earlier can be
  comparable to that of the  \xmm\ with all collected data,
  which will allow one to carry out another independent study of the possible
  sterile neutrino decay signal in this area. In the energy range
  below $\approx2.4$~keV, the expected constraints on the model
  parameters can be significantly
  stronger than those obtained with \xmm. From the \artxc\ data, in the
  energy range approximately from 5 to 20~keV, it can be possible
  to get more stringent constraints than those obtained with \nustar\
  so far. We conclude that the \srg\ mission has a very high potential
  in testing the sterile neutrino dark matter hypothesis.
  
\end{abstract}


\maketitle
\section{Introduction}
\label{sec:intro}

The recently launched Spectrum-Roentgen-Gamma space observatory
(Spektr-RG or \srg, \cite{srg2020}), carrying two X-ray telescopes on
board, \emph{Mikhail Pavlinsky} \artxc\
\cite{2011SPIE.8147E..06P,Pavlinskypart1,
  Pavlinskypart2,2019ExA....48..233P,artxc2020} and \erosita\
\cite{2010SPIE.7732E..0UP,2012arXiv1209.3114M,erosita2020}, is
expected to contribute considerably to cosmology by investigating
cosmic large scale structure properties associated with galaxy
clusters
\cite{2009SPIE.7437E..08P,2010SPIE.7732E..0UP,2012arXiv1209.3114M,2018MNRAS.481..613P},
providing a major improvement in cosmological constraints as compared
to the results from earlier X-ray, Sunyaev-Zeldovich and optical
galaxy cluster surveys, e.g.,
\cite{2009ApJ...692.1060V,2016AA...594A..24P,2020arXiv200211124D}.
While these studies can further refine the parameters of Standard
Cosmological Model (presently, $\Lambda$CDM), the \srg\ has also high
potential in testing specific particle physics models of dark energy
and dark matter. Notably, dark matter particles decaying or
annihilating with keV-scale photons in the final state can be probed
with \erosita\ (energy range 0.2--10\,keV) and \artxc\ (4--30\,keV)
X-ray telescopes aboard \srg.

In this paper we concentrate on a particular candidate of dark 
matter---sterile neutrinos---unstable because of mixing with active
neutrinos and consequently exhibiting a two-body radiative decay into
active neutrino (electron, muon or tau neutrino) and photon
\begin{equation}
  \label{decays}
\nu_{s} \rightarrow \nu_{e, \mu, \tau} + \gamma
\end{equation}
with decay rate\,\cite {PhysRevD.25.766,Barger:1995ty}
\begin{align}\label{eq:neutrino_width}
\nonumber\Gamma_{\gamma} &= \frac{9}{1024} \frac{\alpha}{\pi^4} G_F^2 m_s^5\sin^22\theta\\
&= 1.36 \times 10^{-22}\left(\frac{m_s}{1 \text{keV}}\right)^5 \sin^22\theta\hspace{0.25cm} \text{s}^{-1},
\end{align}
where $m_s$ is the sterile neutrino mass, and $\theta$ is the
sterile-active mixing angle; for small mixing there is
no need to distinguish between the mass and flavor states of the
sterile fermion. The outgoing photon energy is $E_{\gamma}
=m_s/2$, and for sterile neutrinos forming galactic dark matter the
expected photon spectrum is highly monochromatic with width of order
the velocity of the dark matter particles in the galaxy, i.e. $v\sim
10^{-4}\!-\!10^{-3}$. This suggests a signature in the galactic photon
spectrum \cite{Abazajian:2001vt} to be searched for by the X-ray telescopes.

\begin{figure}[!htb]
	\centering
	\includegraphics[width=1.0\linewidth]{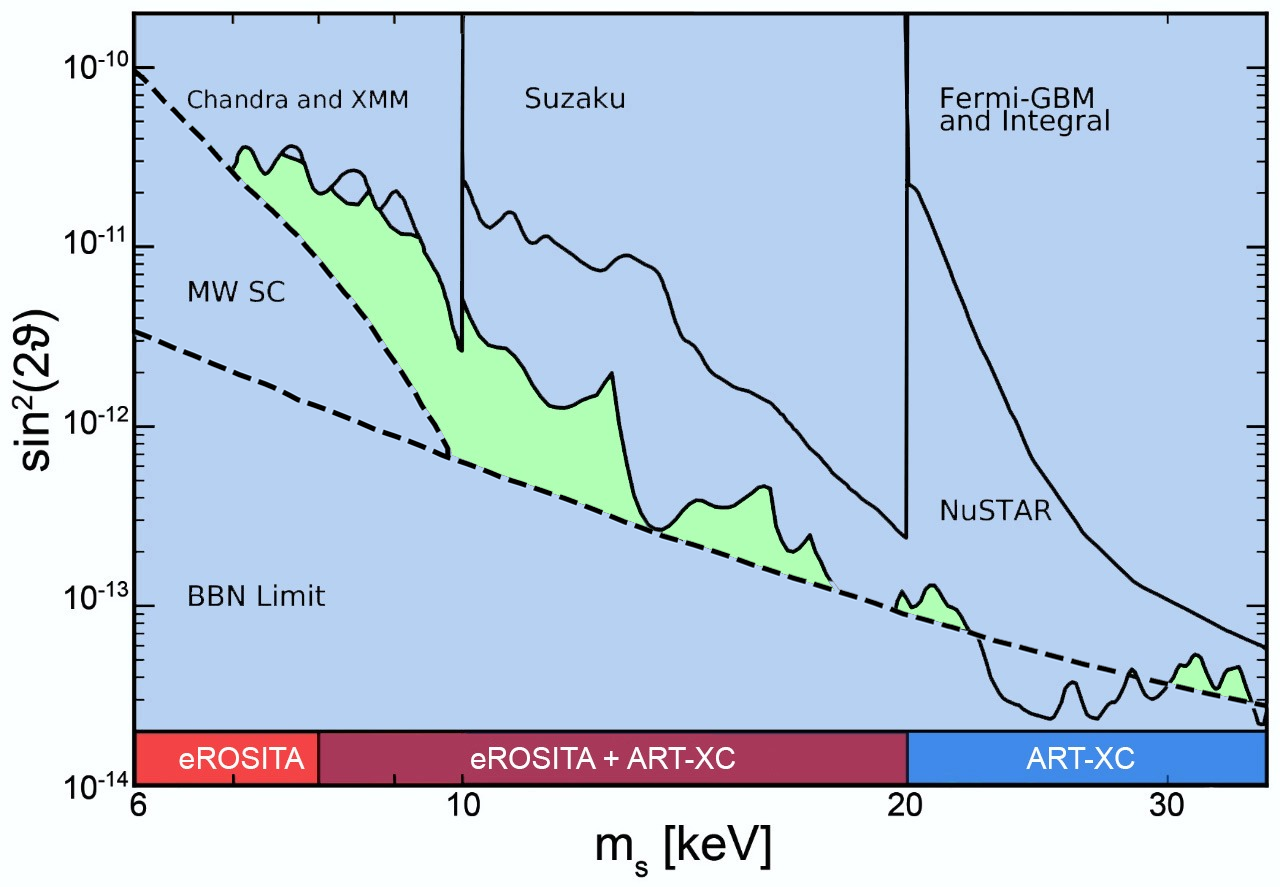}
	\caption{Constraints on the parameters of sterile neutrinos,  
           the summary plot is adopted  
          from Ref.~\cite{Roach:2019ctw}. The green color
          indicates the allowed region of sterile neutrino parameter 
          space consistent with X-ray searches  (solid lines
          indicate the upper limits), cosmological limits
          from Big Bang Nucleosynthesis (dashed line refers to
          the lower bounds) and cosmological constraints from MW
          satellite galaxy counts (dashed line). The X-ray bounds are
      weaker if  the sterile neutrinos form only a subdominant
      component of dark matter. The limits from galaxy counts largely
      depend 
      also on the sterile neutrino spectrum, and hence on its
      production mechanism operating in the early Universe. The BBN
      limits may be relaxed in an extended model, e.g.
      with more ingredients in the neutrino
      sector. The red color shows the energy range
          ($E_{\gamma} = m_s / 2$) covered by the \erosita\ telescope;
          the blue shows the energy range covered by the \artxc\
          telescope; the purple refers the overlapping region of the
          energy ranges of these two telescopes.}
	\label{fig:constraints}
\end{figure}

In the simplest models the sterile neutrinos are Majorana fermions,
singlets with respect to the gauge group of the Standard Model of
particle physics (SM). The introduction of sterile neutrinos into
particle physics is motivated primarily by their contribution to the
active neutrino masses coming (after electroweak symmetry breaking)
from Yukawa-type interactions with the SM Higgs doublet and lepton
doublets. To provide all the three active neutrinos with masses, one
needs at least three sterile neutrinos, but the lightest (or fourth,
fifth, etc) can be sufficiently feebly coupled and hence long-lived to
form dark matter, see Refs.\,\cite{Adhikari:2016bei,Boyarsky:2018tvu}
for a review. Naturally, this strong physical motivation makes the
model very attractive phenomenologically \cite{Abazajian:2012ys} and
the suggested signature has been exploited by X-ray telescopes to
explore the relevant part of model parameter space $(m_s,\theta)$. The
absence of monochromatic photons of a given frequency implies an upper
limit on the mixing angle at the sterile neutrino mass equal to the
double photon frequency provided by
eq.\,\eqref{eq:neutrino_width}. The latest results are provided
  by the \nustar\ experiment, see Fig.\,\ref{fig:constraints} for
  relevant astrophysical and cosmological constraints. 

As a component being in thermal equilibrium with primordial plasma in
the early Universe, the sterile neutrino making all dark matter is not suitable,
because of the same (qualitatively) problems with cosmic structure
formation as active neutrinos have. However, they can be produced
non-thermally by active neutrinos oscillating in the primordial
plasma, that works with relatively small mixing. Note, that the
simplest case, where the mixing is in one-to-one correspondence with
the relic sterile neutrino abundance \cite{Dodelson:1993je}, has been
already excluded\,\cite{Boyarsky:2018tvu}. Nevertheless, for the efficient production of
sterile neutrinos in the early Universe a much smaller mixing (below
upper X-ray limits on \eqref{eq:neutrino_width}) is still sufficient
in cosmological models with lepton asymmetry in the primordial
plasma\,\cite{Shi:1998km} or physical model extensions, see, e.g.,
Refs.\,\cite{Shaposhnikov:2006xi,Bezrukov:2014nza,Bezrukov:2018wvd}
for scalars coupled to sterile neutrinos. Along with particle physics
motivations, these model extensions are supported by a feature at near
3.5\,keV, that claimed to be observed in spectra of several dark
matter dominated astrophysical objects 
\cite{Bulbul:2014sua,Boyarsky:2014jta} (see, however,
\cite{2015MNRAS.452.3905A,2017ApJ...837L..15A,2020Sci...367.1465D}). Therefore,
hunts for a monochromatic line in cosmic X-rays, presumably initiated
by decays\,\eqref{decays}, remain worthwhile and promising in regards to
revealing the nature of dark matter. In this paper we estimate the
sensitivity of \srg\ telescopes to the models with sterile neutrino
dark matter.

If dark matter consists of the decaying sterile neutrinos, then we
expect to observe correlated signal fluxes from various dark matter
dominated astronomical objects. In this paper, we consider the
Galactic Center (GC), the Andromeda Galaxy (M31), and the Draco dwarf
Galaxy (DdG) as possible sources of the monochromatic photons.  This
choice is motivated by investigations based on analyses of
previous generation X-ray telescopes, which found them as the most
prospective sources of the monochromatic line and scrutinized their
dark matter structure.

\section{Expected signal}

\begin{figure}[!htb]
	\centering
	\includegraphics[width=1.0\linewidth]{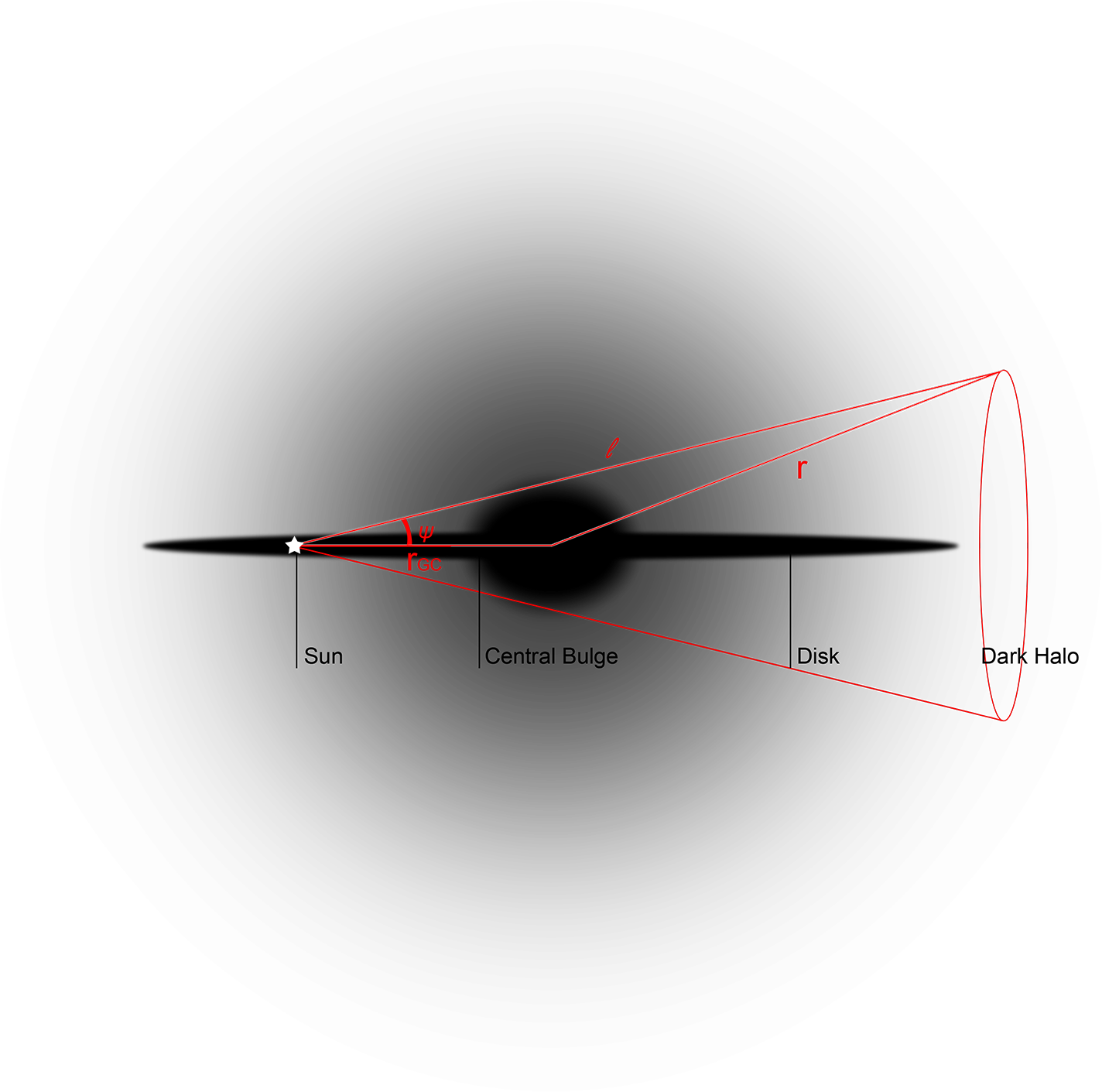}
	\caption{Sketch illustrating the signal geometry, when the Galaxy halo is observed.}
	\label{fig:milkywayprofile}
\end{figure}

Assuming that the sterile neutrinos form (a part of) galactic dark
matter,   
we can estimate the photon flux from decays\,\eqref{decays} in a
nearby source (e.g. Galactic Center, a Milky Way satellite, a Local Group member) as follows, 
\begin{equation}\label{eq:gamma_flux_th}
F_{DM} = \frac{1}{4\pi}
\frac{\Gamma_{\gamma}}{m_s} \mathcal{S}_{DM}\,.
\end{equation}
Here $\mathcal{S}_{DM}$ is the sterile neutrino dark matter column
density of the observed object along the line of sight in the given circular field of view (FOV) towards the center of dark halo of the object, 
\begin{equation}
\label{eq:coldens}
\mathcal{S}_{DM} = \int_{0}^{2\pi}\int_{0}^{w_r}\int_{l_{min}}^{l_{max}}\frac{\rho_{DM}(r(l, \psi))}{l^2}l^2\sin\psi\text{d}\phi\text{d}\psi\text{d}l,
\end{equation}
where $w_r$ is the FOV  
radius of telescope, $\rho_{DM}(r(l, \psi))$ is the dark matter
density profile, and $r(l, \psi) = \sqrt{R^2 -
	2R l\cos\psi + l^2} $, $\psi$ is the angle
  between the line of sight and direction to the center of the object, $ R $ is the distance from the observer to the center of the dark matter halo, e.g.,  distance to the Galaxy center in the case of Milky Way, $l$ is the distance along the line of sight\footnote{For MW we set $l_{min} = 0$ and
  $l_{max} \approx R + R_{vir}$ and for other far objects
  we set $l_{min} \approx R- R_{vir}$ and $l_{max}
  \approx R + R_{vir}$. Multiplying  the virial radius of
  the object   
  $R_{vir}$ by factor of two in these formulas changes the expected
  flux by less than one per cent for M31 and about ten per cent for DdG.}, see
Fig.~\ref{fig:milkywayprofile}.
Thus, the photon flux from the sterile
neutrino decays \eqref{decays} is given by the following expression 
\begin{align}
\label{eq:normalize_flux}
  \nonumber F_{\text{DM}} = &\frac{1}{7.88 \times 10^{-4}} \left(\frac{S_{DM}}{\text{M}_{\bigodot}\text{pc}^{-2}}\right)\\
&\times \left(\frac{2E_{\gamma}}{1\,\text{keV}}\right)^4\sin^22\theta \hspace{0.25cm} \frac{\text{cts}}{\text{$\text{cm}^2$ s}}.
\end{align}

\begin{figure}[!htb]
	\centering
	\includegraphics[width=1.0\linewidth]{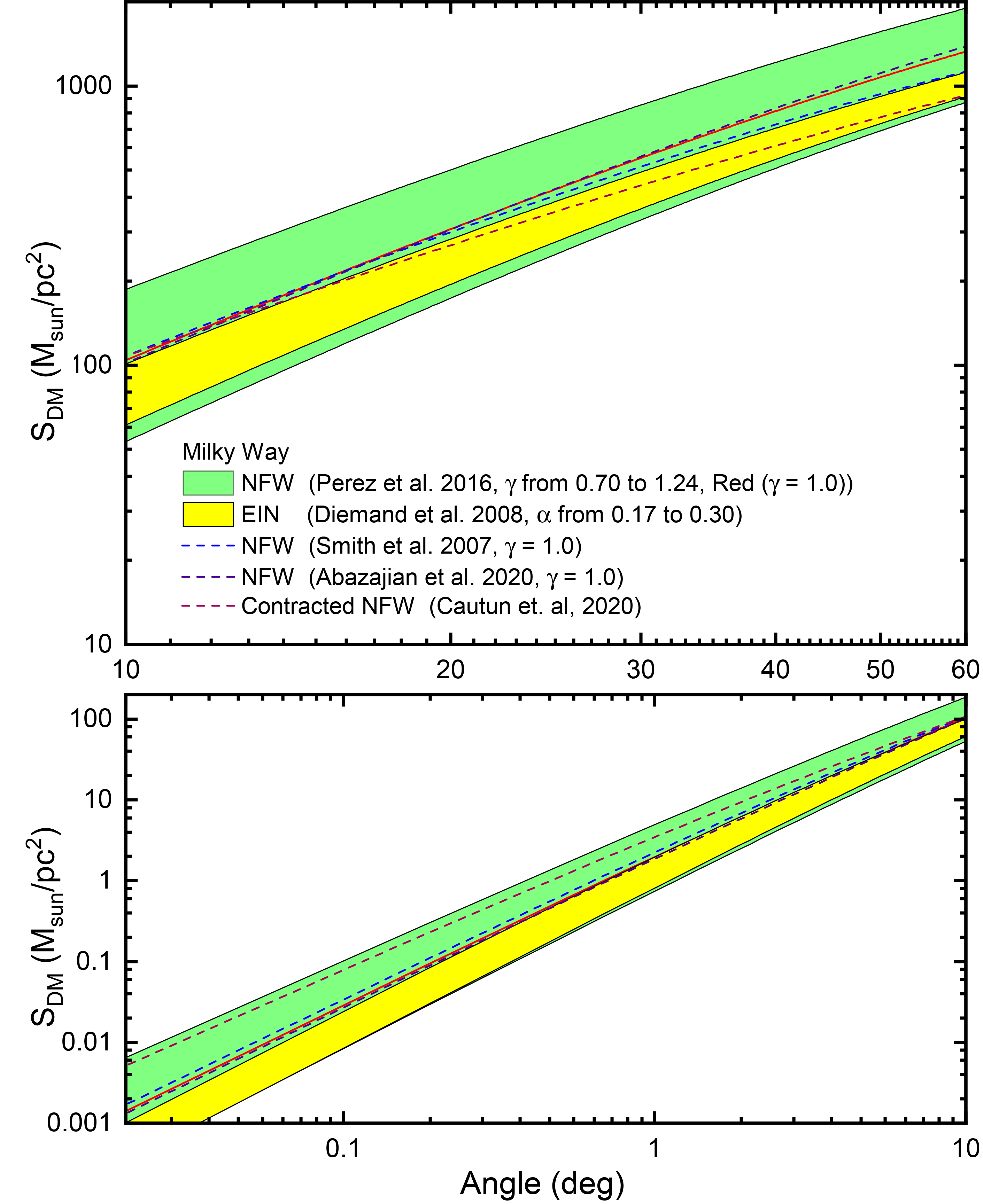}
	\caption{Variance in the DM signal flux at a given cut on the
          viewing angle $\theta$ (towards the MW center) due to difference in the DM profile
          estimates obtained in Refs.\,\cite{Smith:2006ym,Diemand:2008in,2017PhRvD..95l3002P,Cautun:2019eaf,Abazajian:2020tww}.}
	\label{fig:MW-var}
\end{figure}

To evaluate the signal flux, we need to know the distribution density
of dark matter for each object, which is the source of the photons, under
study. This quantity is neither fixed from the theory, nor directly
measured, and so has many uncertainties. However, they are not so
dramatic for the decay signal, as compared to the annihilation of the
dark matter particles, which signal is proportional to squared dark
matter density. In our study, for MW and for M31, we use the standard
NFW profile: $\rho(r) = \rho_{s}/\left(r/r_s\right)\left(1 +
r/r_s\right)^{2}$ \cite{Navarro:1995iw}. For MW, we use the parameters
given in Ref.~\cite{2017PhRvD..95l3002P}, where the values of the scale
density and scale radius are $\rho_{s} = 10.5\times
10^{-3}\text{M}_{\bigodot} \text{pc}^{-3}$ and $r_s = 20$\,kpc, respectively, $R_{GC}=8$\,kpc and
$R_{vir}=200$\,kpc\,\cite{Dehnen:2006cm}. In Fig.\,\ref{fig:MW-var}
we
illustrate the associated with variance of the DM profile
uncertainties in the geometrical factor ${\cal
  S}_{DM}$ entering formula for the expected signal flux
\eqref{eq:gamma_flux_th}. At the very center of MW (small angles) the differences in
the suggested in literature MW profiles are rather dramatic. However,
below we are interested in angles exceeding 
one degree, where the presented uncertainties are at the level
of factor two, which we keep in mind.   
For the M31 halo, we use
the parameter values from~\cite{2012AA...546A...4T}, where $\rho_{s} =
11.0\times 10^{-3}\text{M}_{\bigodot} \text{pc}^{-3}$ and $r_s =
16.5$\,kpc, $R_{GC}=780$\,kpc and
$R_{vir}=207$\,kpc. The matter density distribution
in Draco galaxy is well described by $\rho(r) = \rho_{s}/\left(r/r_s\right)^{\gamma}\left(1 + (r/r_s)^{\alpha}\right)^{(\beta - \gamma) / \alpha}$, with
parameters $\rho_{s} = 18.2\times
10^{-3}\text{M}_{\bigodot} \text{pc}^{-3}$, $r_s = 3.72$\,kpc,
and ($\alpha$, $\beta$, $\gamma$) = ($2.01$, $6.34$,
$0.71$) and $R_{GC}=76$\,kpc, $R_{vir}=1.87$\,kpc \cite{Geringer-Sameth:2014yza}.
Similar to the MW case, the different estimates of the M31 and Draco
DM profiles, see e.g. Refs.\,\cite{Karwin:2019jpy,2009arXiv0901.2569R,
  2019MNRAS.482.3480P}, imply a factor of two uncertainties in the
predictions for the signal flux.

Using the latest constraints on the parameters of sterile neutrinos
presented in \cite{Roach:2019ctw, Ng:2019gch}, we estimate the
possible signal flux from these objects, and
also evaluate the necessary time of observation of these objects by
\srg\ to strengthen the existing constraints, and to check the dark
matter explanation of the 3.5\,keV line.


\section{\srg\ X-ray telescopes}
\label{sec:srg}

The Spectrum-Roentgen-Gamma\,\cite{srg2020} is a Russian X-ray
observatory created with participation of Germany, launched in July,
2019 and designed to produce a deep X-ray map of the Universe in a wide,
0.2--30~keV, energy range. Its scientific payload consists of two
X-ray telescopes --- \artxc\ \cite{2011SPIE.8147E..06P,Pavlinskypart1,Pavlinskypart2,2019ExA....48..233P,artxc2020} and \erosita\
\cite{2010SPIE.7732E..0UP,2012arXiv1209.3114M,erosita2020} --- built in Russia and
Germany, respectively. A map of the large scale structure of the
Universe including more than a hundred thousand clusters of galaxies
will be outlined. At the end of the main survey mission (four years of
operation), the \srg\ will be switched to the pointed observations
of selected objects (two years), which will allow one to explore the
most interesting X-ray sources in more detail. In particular, using
these data, it will be possible to use the spectra of various
astrophysical objects to search for the peak signature of dark matter
radiative decays.

The \srg\ observatory revolves with a six-month period around the second
Lagrange point (L2) of the Sun--Earth system, located at the
distance of approximately 1.5 million km from Earth, in an
elliptical non-closed orbit with 0.75 million km and 0.25 million km
semi-axes. In the survey mode the telescope pointing axis is 
continuously rotated approximately around the direction to Earth
and Sun, therefore, due to this observing strategy the entire
celestial sphere is covered in a six-month period, and eight
full scans of the celestial sphere will be performed during the main four
year stage of operation. The full field of view of the
\erosita\ telescope is 0.8 deg$^2$ and for \artxc\ is around 2 deg$^2$
(in telescope+concentrator mode, see \cite{2019ExA....48..233P}). This
means that the total average exposure time for any part of the sky
will be about 2500 s and 6100 s (uncorrected for vignetting). Note
that by the end of the survey mission, the largest exposure is
expected in the region of the ecliptic poles, since for each
revolution the telescope axis passes near these poles.

\begin{figure}
  \centering
  \includegraphics[width=1.0\linewidth,viewport=7 150 578 690]{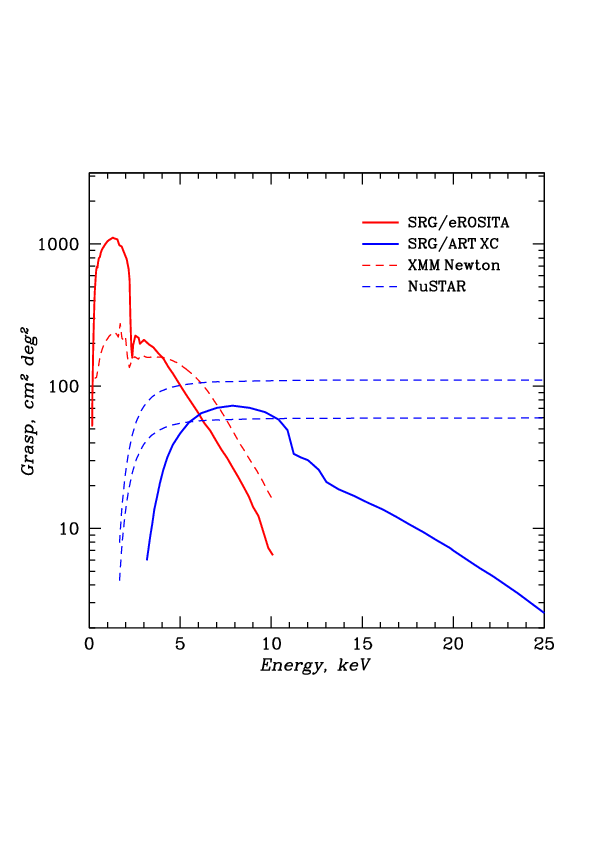}
  \caption{Grasps for \srg/\erosita, \srg/\artxc, \xmm\ and \nustar\
    telescopes, available for sterile neutrino DM decay searches. For
    \srg/\artxc\ telescope grasp for telescope+concentrator mode is
    shown. Lower and upper \nustar\ grasp curves are calculated for
    stray-light aperture only, as used in DM search studies in the GC
    \cite{2017PhRvD..95l3002P} and Bulge \cite{Roach:2019ctw}
    respectively (see text for details).}
  \label{fig:grasp}
\end{figure}

In the survey mode, the most important parameter characterizing the
ability to cover a large part of the sky is ``grasp'', which is
defined as the product of the effective area of the telescope and the
angular area of its field of view corrected for vignetting. Grasps for
\erosita\ and \artxc\ telescopes on board \srg\ observatory, available
for investigations of the sterile neutrino dark matter radiative decay signatures, are shown
in Fig.~\ref{fig:grasp} in comparison with those of \xmm\
\cite{2001AA...365L...1J} and \nustar\ \cite{2013ApJ...770..103H}
telescopes. The grasp of the \artxc\ telescope is shown for
telescope+concentrator mode.  The grasp for \xmm\ is
calculated using PN, MOS1 and MOS2 effective areas and vignetting taken
from \xmm\ Users Handbook. The grasp for the \nustar\ telescope
is calculated for stray-light aperture mode, used in 
the sterile neutrino DM decay searches
\cite{2017PhRvD..95l3002P,Roach:2019ctw}. The lower and upper curves represent average \nustar\ side aperture
(or stray-light) grasp inferred from the \nustar\ observations in the
GC \cite{2017PhRvD..95l3002P} and Bulge \cite{Roach:2019ctw}, where
the most recent DM decay constraints were achieved. The average
\nustar\ grasp is multiplied by the energy-dependent efficiency for
photons to pass through the \nustar\ detector beryllium shield. The
difference in the \nustar\ grasp estimations is mainly explained by a
strong stray-light contamination of bright sources in the GC. In the
Galactic Bulge observations, \nustar\ stray-light aperture
contamination from the X-ray background is minimized, however the
expected DM signal is also lower compared to that from the GC due to
the smaller amount of DM integrated over the line of sight. Note that
the total effective exposure  used for the \nustar\ DM studies in
the GC \cite{2017PhRvD..95l3002P} and Bulge \cite{Roach:2019ctw} is
$\sim$200 and $\sim$100\,ks respectively. At larger angular distances
from GC the search for sterile neutrino DM decay signal was performed with
higher \nustar\ exposure (7.5 Msec) and comparable upper limits were obtained
\cite{Neronov:2016wdd}.

From Fig.~\ref{fig:grasp} we see that the grasp of the \erosita\
telescope significantly supersedes that of \xmm\ at the energies below
$\approx 2.2$~keV. At higher energies the grasps of the telescopes
aboard \srg\ are comparable to those of \xmm\ and \nustar. Therefore,
the main difference in DM decay flux constraints at these energies
between \erosita\ and \artxc\ telescopes as compared to
\xmm\ and \nustar, will arise due to the difference in exposures used
to observe the source. As we show below, when 
the dark matter halo of our Galaxy is observed, with the \srg\ telescopes it will
be possible to collect the data with significantly higher exposures,
as compared to what were used to constrain the sterile neutrino DM
from \xmm\ and \nustar\ data.

For the reliable signal detection, it is necessary to know energy
dependence of the background flux. Estimates of the expected number of
events for the \erosita\ and \artxc\ telescopes obtained by modeling
the background signal and ground calibrations, as well as based on
observational data from earlier missions are presented in
\cite{2012arXiv1209.3114M,2019ExA....48..233P}. The charged particles
induced background, observed during the first months of the \srg\ flight is
found to be in general agreement with preflight estimates for \artxc\
telescope \cite{srg2020,artxc2020} and a few times higher for \erosita\
\cite{srg2020,erosita2020}.  The particle background should become
lower, when Sun activity will be higher
(e.g., \cite{Gonzalez-Riestra2019}), which should happen during the
next few years.

For the estimates below we adopt the observed particle background of
\erosita\ and \artxc\ taken from
Refs.~\cite{srg2020,artxc2020,erosita2020}. Since \srg\ is placed to
orbit around the second Lagrange point (L2) of the Earth-Sun system, the
thermal conditions onboard \srg\ and particle background on the
detectors of both \artxc\ and \erosita\ telescopes appear to be
extremely stable, as compared to those of missions at the Earth
orbits.

At the energies below 2~keV, the background of the \erosita\ telescope
is dominated by Cosmic X-ray Background (CXB) all over the sky and,
additionally, by the Galactic Ridge X-ray Emission (GRXE)
\cite{revnivtsev2006,krivonos2007,2009Natur.458.1142R,2019ApJ...884..153P}
when pointed towards the Galactic plane and bulge. We expect that
approximately 50\% of CXB will be resolved by \erosita\ at its survey
limiting flux of $\sim10^{-14}$~erg~s$^{-1}$~cm$^{-2}$, 0.5--2~keV,
see e.g. \cite{2002A&A...389...93L,2003ApJ...588..696M}. The detected X-ray sources will be removed from the data. Due to the good angular resolution of eROSITA and ART-XC, the loss of effective FOV in the extragalactic fields will not be significant, much less than 1\%, which can be easily estimated given the angular resolution of the telescopes and the number of expected source detections. The loss of effective FOV due to the exclusion of bright X-ray sources near the Galactic Center will be larger, up to a few percents, which will still  not be significant for our purposes. For weaker sources with flux below ${\sim}$mCrab, we calculated the contribution from the GRXE to the background model based on the near-infrared stellar mass model from Ref.\,\cite{launhardt2002} and 3--20~keV Galactic ridge X-ray emissivity per unit stellar mass from Ref.\,\cite{revnivtsev2006}.  This approach has been proven to effectively model the observed number-flux distribution of detected sources in the Chandra deep survey of the Norma spiral arm region \citep{2014ApJ...796..105F}. Note that, when analysing the actual SRG data, the higher number of detected Galactic X-ray sources will lead, on the one side, to larger FOV losses, and on the other side, to decreased GRXE contribution to the background model.

The systematic uncertainties due to the background modeling are
neglected here. We note however that to obtain the constraints discussed below the systematic uncertainties should be controlled at the level of $\sim1\%$. We expect that these uncertainties in SRG telescopes data will be lower as compared to those in previous experiments, because the observed particle background is found to be very stable and it will be possible to obtain direct and accurate background measurements for both telescopes. Further details of SRG telescopes background modeling will be discussed in our future papers.

We also assume that the sterile neutrino DM decay flux is detected in
a bandwidth equal to the energy resolution of the corresponding
telescope, which are taken from \cite{2012arXiv1209.3114M} for
\srg/\erosita\ and from \cite{Pavlinskypart1} for \srg/\artxc\ and are
summarized in Table~\ref{tab:ttp}.

\begin{table}[!htb]
  \caption{Telescopes technical performance.}
  \label{tab:ttp}
  \begin{center}
    \begin{tabular}{lcc}
      \hline
      \hline
      {} & \erosita\ & \artxc\ \\
      \hline
      energy range [keV] & 0.2 -- 10 & 4 -- 30\\
      energy resolution, FWHM & 138\,eV at 6\,keV & 10\% at 14\,keV \\
      field of view (FOV) [deg$^2$] & 0.833 & 0.3 -- 2.0 \footnote{FOV [deg$^2$]: Telescope 0.31, Concentrator 1.7, Full 2.0 \cite{Pavlinskypart1, Pavlinskypart2, 2019ExA....48..233P}.}\\
      \hline
    \end{tabular}
  \end{center}
\end{table}

\section{Results: expected constraints on the Dark Matter decay flux}
\label{sec:constraints}

The existing constraints on the parameters of sterile neutrinos,
presented in Ref.~\cite{Roach:2019ctw}, can be used to estimate the
ranges of possible signal flux from decays of resonantly produced dark
matter sterile neutrino to be observed by the
\srg\ mission. For this purpose, we calculate the photon
  flux (\ref{eq:normalize_flux}) from decays of sterile neutrinos in
  the central part of MW, M31 and Draco. The corresponding sterile
  neutrino dark matter column density factors for MW, M31 and Draco
  for particular circular region radii in the formula for signal flux
  \eqref{eq:normalize_flux} are obtained in this way as (in
  $M_\odot\text{pc}^{-2}$)
\begin{equation}
  \label{col-den}
  \begin{split}
&{S}_{DM}^{\text{GC}_{3.0^\circ}}=13.7,\;\;\;{S}_{DM}^{\text{GC}_{60^\circ}}=1.3\times 10^3\,,\\
&{S}_{DM}^{\text{M31}_{1.5^\circ}}=0.39,\;\;{S}_{DM}^{\text{Draco}_{0.25^\circ}}=1.4\times 10^{-2}.
\end{split}
\end{equation}
Within the minimal variants of the sterile neutrino model we are
  concentrated on, the signal fluxes are constrained from above by
  upper limits on mixing angle $\theta$ (entering
  eq.\,\eqref{eq:normalize_flux}) placed by previous X-ray
  experiments. Likewise the fluxes are constrained from below by
  coming from cosmology lower limits on mixing angle. Hence, for each
  astrophysical source the expected from the dark matter decays flux
  is bounded from above and from below.

\begin{figure}[!htb]
  \includegraphics[width=\columnwidth]{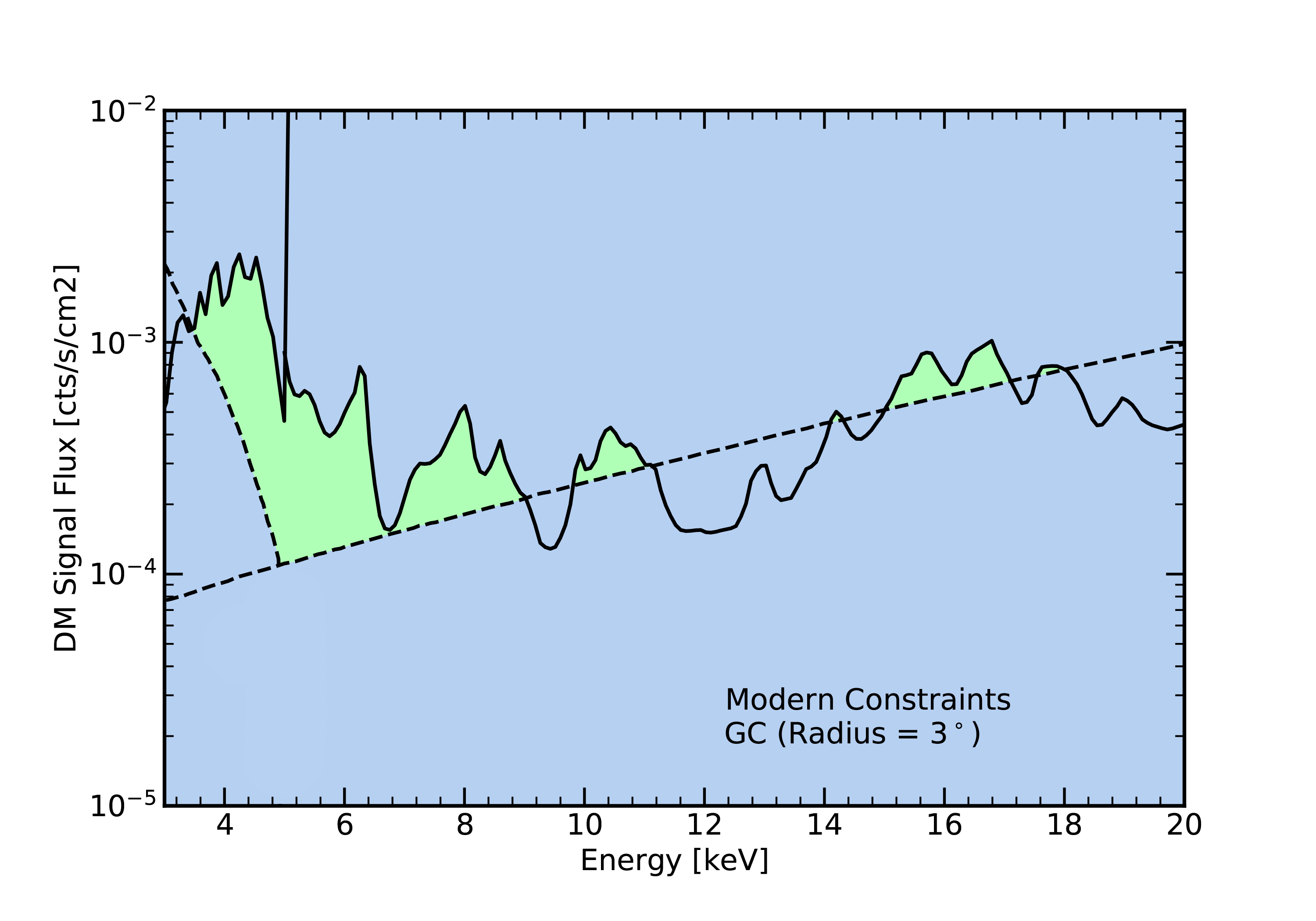}
 
  \caption{The allowed region of the possible signal flux\,\eqref{eq:normalize_flux} (green) from
    decays of dark matter particles, which can be tested with the
    \erosita\ and \artxc\ telescopes on board \srg\ observatory in GC
    pointed observations (the circular region of radius
    $3^\circ$). The regions in blue are already excluded from  X-ray telescopes
    (regions above the solid lines) and cosmology assuming the
    resonant mechanism responsible for the sterile neutrino production
    in the early Universe (regions below the dashed lines).}
  \label{fig:possible_flux}
\end{figure}

With limits depicted in Fig.\,\ref{fig:constraints}, we estimate the   X-ray signal flux\,\eqref{eq:normalize_flux} for each source, assuming background model in Sec.~\ref{sec:srg}. For the \erosita\ observation of   GC with radius of $60^\circ$ we cut the central part of $2.5^\circ$   radius and the Galactic disc of $\pm1.5^\circ$.   Figure~\ref{fig:possible_flux} illustrates the possible signal flux   from observations of the GC: any peak-like feature in the X-ray with parameters (flux and energy) inside the green region may hint at its   dark matter nature within the described sterile neutrino model.  We use the estimated fluxes to evaluate the observation time $T$,   required to strengthen the existing limits (i.e. to enter the green regions similar to that in Fig.\,\ref{fig:possible_flux}). Namely,   observation for a period $T$ allows one to place a limit on photon flux $F_{DM}$ at a confidence level (in terms of the standard deviation $\sigma$) if the expected flux is not detected: 
\begin{equation}
\label{eq:est_time}
T = \left[\frac{F_{DM} \times G(E)/\Omega + C_{BG}}{F^2_{DM} \times G^2(E)/\Omega^2}\right] \sigma^2,
\end{equation}
where $G(E)/\Omega$ is the ratio of grasp to angular size of the source and $C_{BG}$ is the background count rate (in counts per second). Note, that this equation is valid only in case when source angular size is larger than the telescope FOV.  If one observes a region equal to the field of view of
the telescope ($\Omega \cong \text{FOV}$), then this ratio is the
effective area (EA) of the telescope. 
With \srg\
  survey strategy (see Sec.~\ref{sec:srg}), both \artxc\ and
  \erosita\ sky coverage inside the cones centered on GC can be
  considered for our purpose as approximately uniform even for $60^\circ$ cone
  radius.  The count rate of the background events $C_{BG}$ here is taken in a bandwidth corresponding to the
telescope energy resolution. The energy resolution of the \erosita\
telescope remains almost constant, 2.3\%, over the entire observation
energy range \cite{2012arXiv1209.3114M}. The energy dependence of
\artxc\ telescope energy resolution is taken from
\cite{Pavlinskypart1}.

\begin{figure}
  \includegraphics[width=\columnwidth]{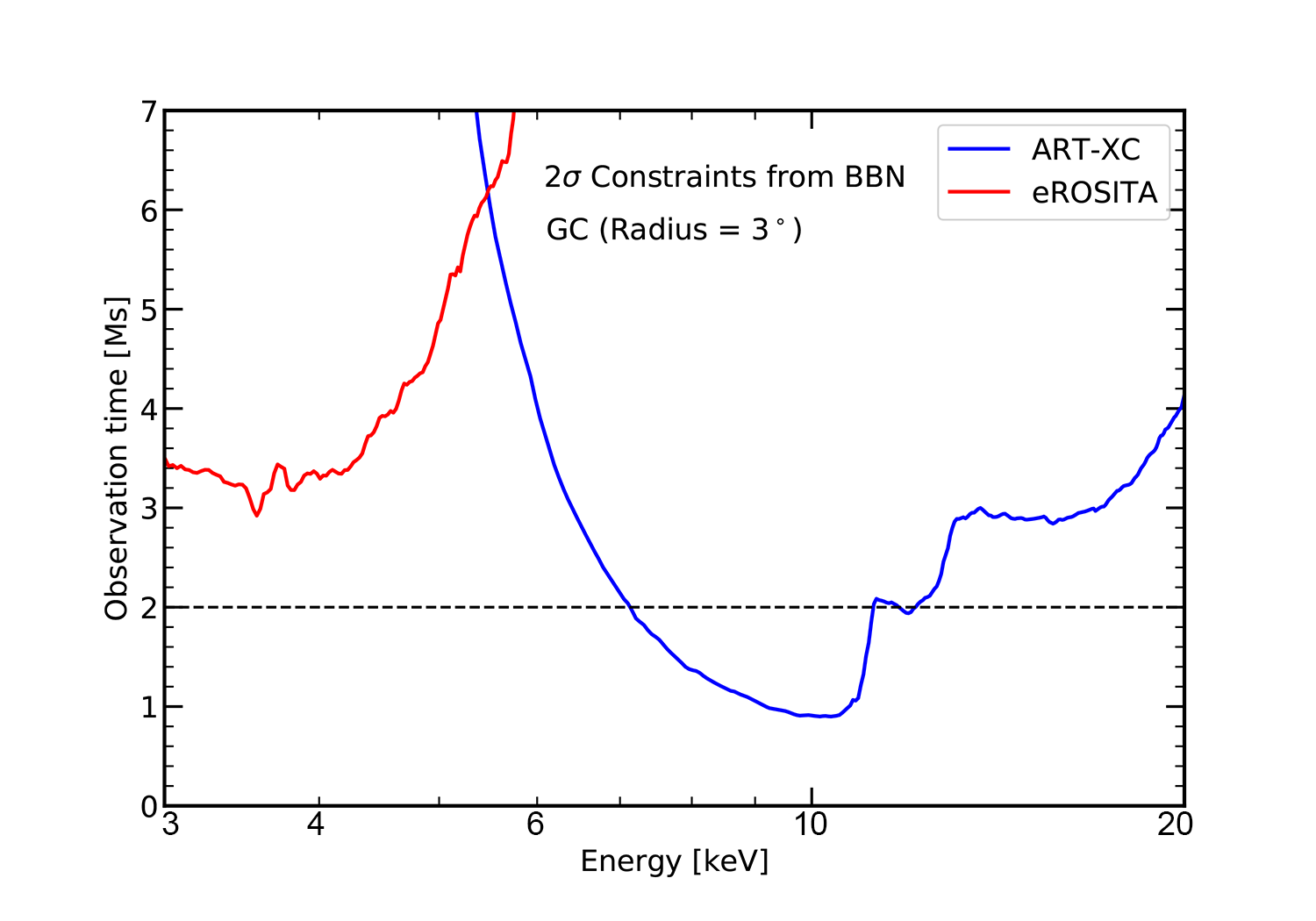}
  \includegraphics[width=\columnwidth]{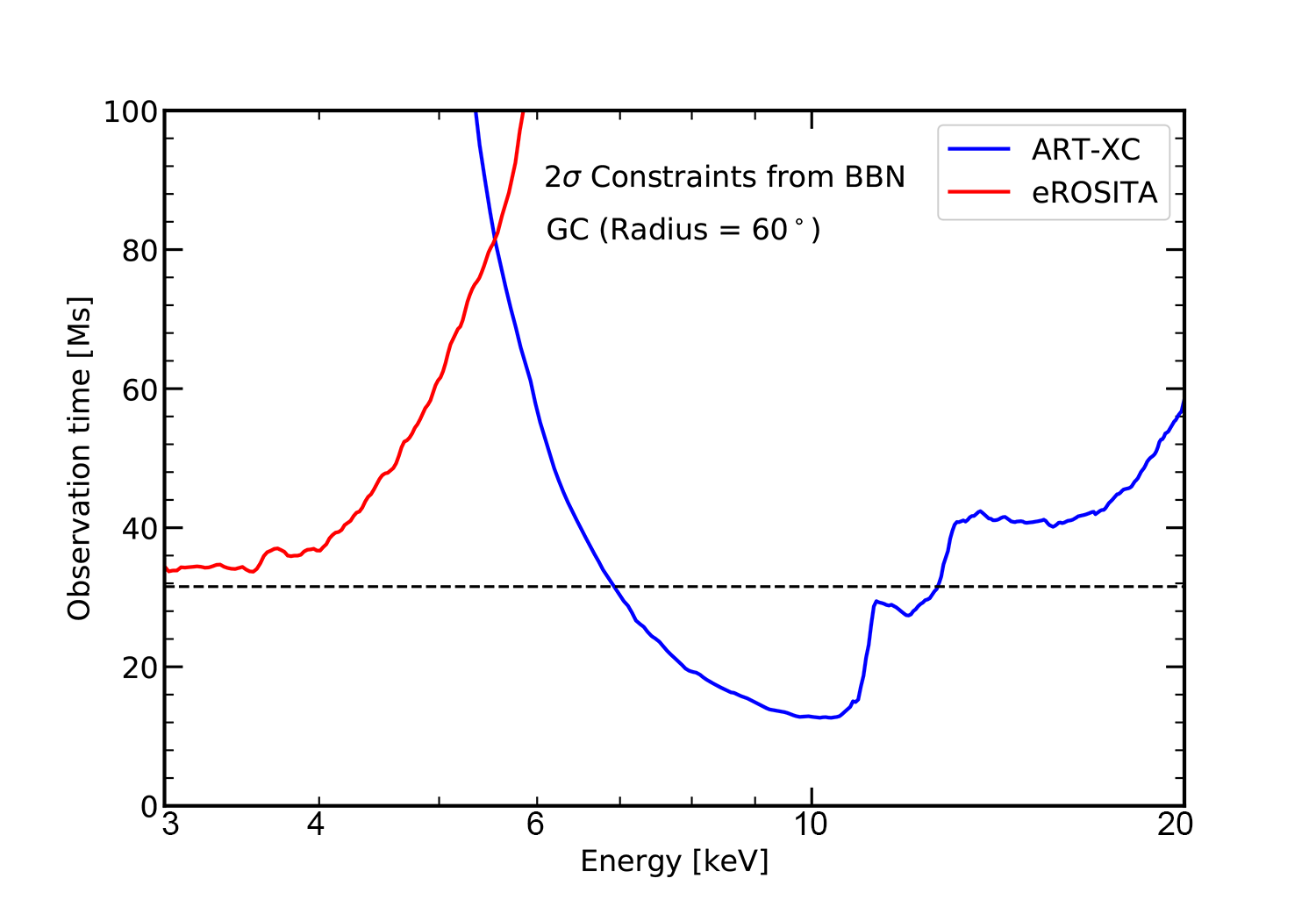}
  
  \caption{The observation time for \erosita\ and \artxc\ telescopes
    for circular regions of radius $3^\circ$ and $60^\circ$ around
    the position of GC required to fully explore the dark matter
    sterile neutrino model with resonant production mechanism
    (exclusion at $2\sigma$ confidence level). The horizontal dashed
    line on the upper panel shows the exposure obtained by
    \artxc\ observation of GC before the start of main survey
    program, the similar line on lower panel shows the exposure to be
    collected in the four-year \srg\ all-sky survey by each of the
    telescopes, \erosita\
    and \artxc.}
  \label{fig:gc_est_time_120}
\end{figure}

Figure~\ref{fig:gc_est_time_120}
shows the estimated exposure time for the GC (in circular regions of
radius $3^\circ$ and $60^\circ$) with the \erosita\ and \artxc\
telescopes. It follows from~\eqref{eq:est_time} for placing
$2\sigma$-limit on the model with the lowest possible mixing sill
yielding the dark matter sterile neutrinos via the resonant production
in the early Universe and consisting with cosmological constraints
from the Big Bang Nucleosynthesis. Horizontal dashed lines in
Fig.~\ref{fig:gc_est_time_120} show the exposure time achievable with
\srg\ telescopes in various sky surveys (see discussion below).

\section{Conclusions and Outlook}
In this paper, we examined the capabilities of the \srg\ mission to
detect a signal from the dark matter formed by sterile neutrinos,
which can decay into active neutrino and photon. We have estimated the
potential signal fluxes which can be detected during the \srg\ mission
from the Galactic Center, the Andromeda and Draco galaxies. For the
Milky Way, we separately examined the expected signal flux for two
different observational strategies: the observations near the Galactic
Center (3$^\circ$ radius towards the center) and in the periphery of
the Milky Way (60$^\circ$ radius around GC). Based on these data, we
estimated the necessary exposure time to observe the Milky Way in
these two cases for a reliable exclusion of the lowest possible signal
from decays of the dark matter sterile neutrinos resonantly produced
in the early Universe.  The modeling of the background spectra
presented in this article does not contain possible systematic
uncertainties. The results we offer provide only an assessment of the
usefulness of \srg\, data in sterile neutrino decay studies. One more
source of uncertainties is associated with variance of the DM
profiles of the chosen halos, which can change the signal flux roughly
by a factor of two. The numerical estimates presented below do 
  not account for this uncertainty.

Remarkably, the estimated exposure time in case of the observations near the
GC is comparable with the exposure of \artxc\ GC survey, which
was carried out in September 2019, shortly after the launch of the
\srg\ 
spacecraft, during the performance and verification observations
phase. In this survey $\approx 40$ square degrees region around GC was 
covered by the \artxc\ for nearly 24 days ($\approx2$~Ms, shown as
the horizontal dashed line on the upper panel of
Fig.~\ref{fig:gc_est_time_120}). Using these data it should be
possible to obtain stronger than 2-$\sigma$ constraints on the lowest
possible dark matter decay signal at the photon energies between
5.5~keV and 19~keV. Unfortunately, only a few \erosita\ detectors were
occasionally switched on in the testing mode during these observations,
so the possibility to obtain strong dark matter decay constraints
using these \erosita\ data should be still investigated. 

The exposure time in a wide 60$^\circ$ radius region towards the 
 GC, is to be compared to the exposure which will be
obtained in the \srg\ all-sky survey. It is currently planned that all the
sky will be surveyed by \srg\ during four years in total. Therefore,
60$^\circ$ radius circular region centered on the GC will be surveyed for
approximately one year during this survey ($\approx 30$~Ms, shown as
the horizontal dashed line on the lower panel of
Fig.~\ref{fig:gc_est_time_120}). These data will allow one to place
stronger than 2-$\sigma$ bounds on the lowest possible dark matter
decay signal at the photon energies between 6.5~keV and 12.5~keV with
\artxc\ and at the photon energies below 6~keV with \erosita. Thus,
our estimates imply that it will be possible to obtain even
stronger constraints on the parameters of sterile neutrinos in the
corresponding mass range using the \artxc\ and \erosita\
all-sky data.

\begin{figure}
  \includegraphics[width=\columnwidth]{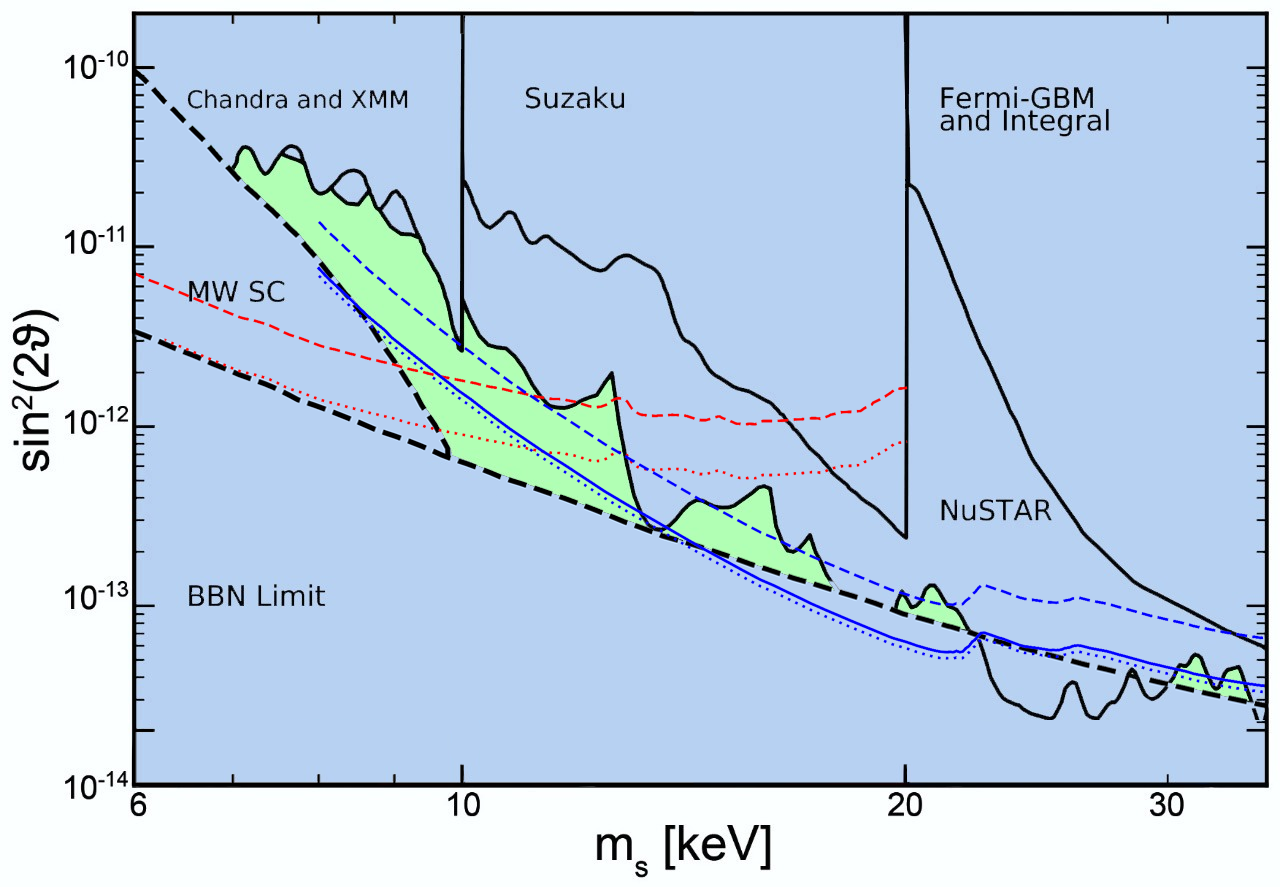}
   \caption{The expected limits on the sterile neutrino model
     parameter space to be obtained (assuming no signal) with already
     collected data towards the GC (solid line) and with
     data to be collected after 1-year (dashed line) and 4-year
     (dotted line) operation in the all-sky survey mode
     and used to analyze the signal 
     from $60^\circ$-region around the GC.  Red (blue) lines refer to
     \erosita\ (\artxc) instruments; black solid and dashed lines
     indicate existing constraints, the same as in
     Fig.\,\ref{fig:constraints}.} 
  \label{fig:new-res}
\end{figure}

To illustrate this conclusion, we present in Fig.\,\ref{fig:new-res}
limits on the mixing angle to be placed by \srg\ (assuming no signal
is observed) with already collected data towards GC  (only
  \artxc, 
  solid blue line), 
and with 1-year  (blue (red) dashed line for \erosita\ (\artxc)) 
and 4-year survey data  (blue (red) dotted blue line for
  \erosita\ (\artxc)) 
considering $60^\circ$ region around GC.  In
particular, to detect at 5-$\sigma$ level the 7\,keV dark matter
sterile neutrino with mixing of about
$\sin^22\theta\sim 5\times 10^{-11}$ (invited to explain the anomalous
3.5\,keV line\,\cite{Bulbul:2014sua,Boyarsky:2014jta}), it takes only
$\sim$ 25~ks and 0.4~Ms of observations near GC and in the wide angle
around GC respectively. The \erosita\ sensitivity at 3.5 keV is
approximately similar to that of \xmm, as expected, see e.g. 
\cite{2020Sci...367.1465D}, also Fig.~\ref{fig:grasp}.  Confronting
these exposures with \srg\ mission time scale one concludes, that the
constraints to be obtained by the \erosita\ telescope near the 3.5 keV
line signal reported earlier can be comparable to that obtained from
\xmm\ data, which will provide with another independent study of the
possible sterile neutrino decay signal in this mass range. 
  Similar and higher sensitivities to the sterile neutrino model are expected for
the next generation X-ray telescopes like Athena and eXTP, see
e.g.\,\cite{Neronov:2015kca,Malyshev:2020hcc}.

At the photon energies below approximately 2.4~keV the \erosita\
sensitivity in wide angle observations is higher than that of
\xmm. Therefore \erosita\ data from a wide circular region around GC
yields a new independent prob of the region of the DM sterile neutrino
parameter space, previously disfavored from the Milky Way satellite
counts \cite{2017PhRvD..95h3015C}, not from X-ray observations. This
may be important, since both methods may have their own unknown
systematics. Likewise, if dark matter particles decay, it happens in
each galaxy and all the time, so the expected X-ray signal from
unresolved astrophysical sources at cosmological distances suffers
from redshifting and correlates with the Cosmic Structure. The
sensitivity of the \erosita\ to the corresponding observables in the
X-ray diffuse all-sky map has been studied in
Refs.\,\cite{Zandanel:2015xca,Caputo:2019djj}. Similar investigation
is envisaged for \artxc. As a kind of intermediate approach, one can
use the stacked spectra of resolved astrophysical sources to search
for the decay signature.

The data of X-ray surveys (full sky coverage) which are carried out
with \artxc\ and \erosita\ telescopes aboard \srg\ space observatory
will give a possibility to fully explore the dark matter sterile
neutrino model with resonant production in the early Universe. The
presence of overlapping areas in the energy ranges of the
\erosita\ and \artxc\ telescopes will further improve the overall
sensitivity to the parameters of sterile neutrinos.  The \srg\ mission
has high potential for verifying the dark matter sterile neutrino
hypothesis within the minimal extensions of the SM like
$\nu$MSM\,\cite{Boyarsky:2009ix}, where no additional ingredients are
introduced in the model to change the neutrino dynamics at the
production epoch. There still be a room for the models with sterile
neutrino forming only a fraction of dark matter.  To conclude, the
\srg\ mission has a very high potential for testing the sterile
neutrino dark matter hypothesis.

\acknowledgements

Authors are grateful to Eugene Churazov, Oleg Ruchayskiy and Mikhail
Shaposhnikov for useful discussion and
important comments. The work on theoretical prediction of DM signal
from decaying sterile neutrinos is supported by the Russian Science
Foundation (RSF) grant 17-12-01547. Calculation of the sensitivity of
X-ray telescopes in wide-angle observations is done under support of
the RSF grant 18-12-00520.

\addcontentsline{toc}{chapter}{\bibname}
\bibliographystyle{apsrev}
\bibliography{refs}

\end{document}